\documentclass[prb,showpacs,floatfix,twocolumn,superscriptaddress,byrevtex]{revtex4}
\usepackage{amsfonts}
\usepackage{dcolumn}
\usepackage{graphicx}
\usepackage{bm}
\begin{document}

\bibliographystyle{apsrev}

\preprint{Draft, not for distribution}
%
%
\title{Infrared phonon anomaly in \boldmath BaFe$_2$As$_2$ \unboldmath }
%
%
%
\author{A. Akrap}
\affiliation{Department of Condensed Matter Physics and Materials Science,%
 Brookhaven National Laboratory, Upton, New York 11973, USA}
\author{J. J. Tu}
\affiliation{Department of Physics, The City College of New York, New York, NY 10031, USA}
\author{L. J. Li}
\author{G. H. Cao}
\author{Z. A. Xu}
\affiliation{Department of Physics, Zhejiang University, Hangzhou 310027, China}
\author{C. C. Homes}
\email{homes@bnl.gov}
\affiliation{Department of Condensed Matter Physics and Materials Science,%
 Brookhaven National Laboratory, Upton, New York 11973, USA}

%
%
\date{\today}

%
%
\begin{abstract}
The detailed optical properties of BaFe$_2$As$_2$ have been determined over a wide
frequency range above and below the structural and magnetic transition at
$T_N \simeq 138$~K.  A prominent in-plane infrared-active mode is observed
at 253~cm$^{-1}$ (31.4~meV) at 295~K.  The frequency of this vibration shifts
discontinuously at $T_N$; for $T < T_N$ the frequency of this mode displays
almost no temperature dependence, yet it nearly doubles in intensity.
This anomalous behavior appears to be a consequence of orbital ordering
in the Fe-As layers.
\end{abstract}
%
%
%
%
\maketitle

%
%
%
The exciting discovery of superconductivity in the iron-arsenic (pnictide) compound
LaFeAsO$_{1-x}$F$_x$ with a high critical temperature\cite{kamihara08} $T_c=26$~K
has generated a great deal of interest in this class of materials. Other rare-earth
substitutions\cite{ren08} quickly increased $T_c$ above 50~K, and $T_c$'s in excess
of 50~K have been also achieved through the application of pressure.\cite{yi08}
More recently, the oxygen free and structurally simpler BaFe$_2$As$_2$ material
has been investigated.  At room  temperature, this material is tetragonal ($I4/mmm)$,
but undergoes a magnetic transition at $T_N \simeq 138$~K that is accompanied at
the same time by a weak structural distortion into an orthorhombic phase ($Fmmm$)
with anomalies in the specific heat, resistivity and susceptibility.\cite{rotter08a,wang09}
While the magnetic transition in the pnictides was originally discussed as a
spin-density-wave instability,\cite{cruz08,rotter08a} there is currently some
debate as to the microscopic nature of the magnetism.\cite{johannes09}
The magnetic and structural transitions are suppressed and superconductivity is
recovered through the application of either pressure\cite{alireza09} ($T_c \simeq 29$~K)
or chemical doping\cite{rotter08b,sasmal08,sefat09} ($T_c=38$~K in the potassium-doped
material), indicating that the superconductivity in this class of materials originates
in the Fe-As layers.
%
%
%
When the relatively high values for $T_c$ are considered with the strong
interplay between the lattice and magnetism, it is likely that the superconducting
pairing interaction is not phonon mediated.\cite{boeri08}
%
%
However, electron-phonon coupling may be present in the pnictides.\cite{eschrig08,kulic09}
Optical investigations of the non-superconducting BaFe$_2$As$_2$ compound\cite{hu08,pfuner09}
and the doped superconducting materials\cite{li08,yang09,tu09} have focused primarily on the
large-scale features in the optical properties; the vibrational features in the undoped
material have either not been observed,\cite{pfuner09} or if they have been observed,\cite{hu08}
they have not been discussed.

%
%
In this report we present the detailed in-plane optical properties of a single
crystal of BaFe$_2$As$_2$.  In addition to the large scale-changes previously
observed in the optical properties,\cite{hu08} we also observe both in-plane
symmetry-allowed infrared-active modes at $\simeq 94$ and 253~cm$^{-1}$
at 295~K.  Anomalous behavior in both the position and strength of the
253~cm$^{-1}$ mode is observed below $T_N$; this mode involves displacements
in the Fe-As layer.  The possible origins of this behavior are discussed, with
the most compelling being an orbital-ordering scenario resulting in a change
in the nature of the bonding.\cite{shimojima09,chen09,phillips09,lee09}

\begin{figure}[tb]
\vspace*{0.1cm}%
\centerline{\includegraphics[width=2.8in]{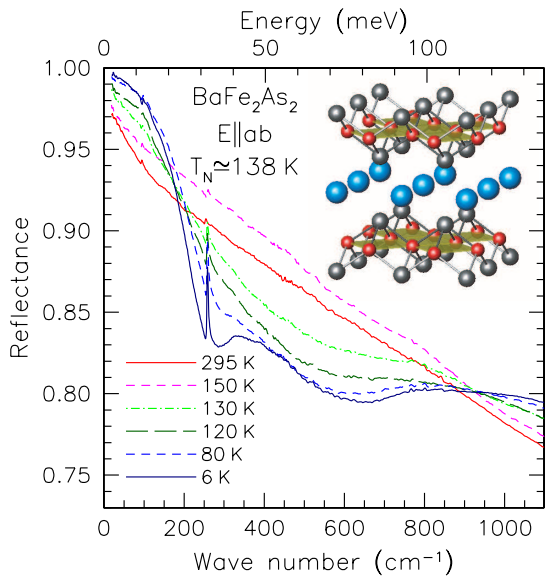}}%
\vspace*{-0.2cm}%
\caption{(Color online).  The reflectance in the low-frequency region for a
single crystal of BaFe$_2$As$_2$ for light polarized in the {\em a-b} planes
at several temperatures above and below the structural and magnetic transition
at $T_N \simeq 138$~K.  The resolution at low frequency is typically better than
2~cm$^{-1}$. Inset: The arrangement of the Fe-As layers and the interstitial Ba
atoms. }
\vspace*{-0.3cm}%
\label{fig:reflec}
\end{figure}

%
%
Large single crystals of BaFe$_2$As$_2$ were grown by a self-flux method.\cite{sefat09}
The reflectance has been measured above and below $T_N$ over a wide frequency range
(2~meV to over 3~eV) for light polarized in the {\em a-b} plane  using an {\em in situ}
evaporation technique.\cite{homes93} The low-frequency results are shown in
Fig.~\ref{fig:reflec}.
%
%
At room temperature the low-frequency reflectance is metallic;
however, there is a prominent shoulder at about 5000~cm$^{-1}$ or 0.6~eV (not shown)
that has been previously observed.\cite{hu08,pfuner09}  As the temperature is reduced
the low-frequency reflectance continues to increase, but for $T < T_N$ the reflectance
between $200 - 800$~cm$^{-1}$ shows a remarkable suppression.\cite{hu08}  We
note that this type of behavior is also observed in chromium\cite{barker70}
below $T_N \simeq 312$~K, and in the charge- and spin-stripe ordered state of
La$_{2-x}$Ba$_x$CuO$_4$ for $x=1/8$ (for which the superconductivity is
dramatically suppressed) where the large changes in the reflectance were
associated with the partial gapping of the Fermi surface.\cite{homes06,valla06}
In addition to the gross features in the reflectance of BaFe$_2$As$_2$, the
two sharp features observed at $\simeq 94$ and 253~cm$^{-1}$ (11.7 and 31.4~meV,
respectively) are the in-plane infrared-active lattice vibrations.

%
%

\begin{figure}[b]
\vspace*{-0.3cm}%
\centerline{\includegraphics[width=2.8in]{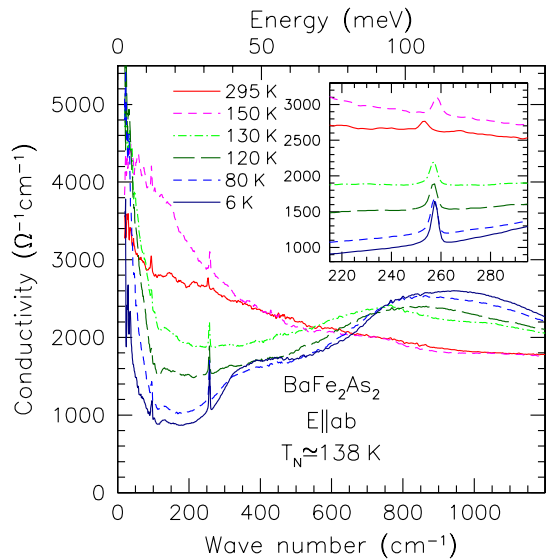}}%
\vspace*{-0.1cm}%
\caption{(Color online). The real part of the optical conductivity in the
  low-frequency region for BaFe$_2$As$_2$ for light polarized in the
  {\it a-b} planes for several temperatures above and below the structural
  and magnetic transition at $T_N \simeq 138$~K.
  Inset: The real part of the optical conductivity in the region of the
  infrared-active mode at $\simeq 253$~cm$^{-1}$. }%
\vspace*{-0.1cm}%
\label{fig:sigma}
\end{figure}

%
%
The optical conductivity has been determined from a Kramers-Kronig analysis of
the reflectance.
%
%
The calculated conductivity is shown in the low-frequency region in Fig.~\ref{fig:sigma}.
The optical conductivity can be modeled reasonably well by using a Drude-Lorentz model
for the complex dielectric function
\begin{equation}
  \tilde\epsilon(\omega) = \epsilon_\infty - {{\omega_{p,D}^2}\over{\omega^2+i\omega/\tau_D}}
    + \sum_j {{\Omega_j^2}\over{\omega_j^2 - \omega^2 - i\omega\gamma_j}},
\end{equation}
where $\epsilon_\infty$ is the real part of the dielectric function at high
frequency, $\omega_{p,D}^2 = 4\pi ne^2/m^\ast$ and $1/\tau_D$ are the plasma frequency
and scattering rate for the delocalized (Drude) carriers, respectively; $\omega_j$,
$\gamma_j$ and $\Omega_j$ are the position, width, and oscillator strength of
the $j$th vibration (the intensity is proportional to $\Omega_j^2$).  The complex
conductivity is simply $\tilde\sigma(\omega) = \sigma_1 +i\sigma_2 =
-i\omega [\tilde\epsilon(\omega) - \epsilon_\infty ]/4\pi$.

Above $T_N$ the conductivity may be reproduced below 1~eV by using a Drude term
in combination with several bound excitations; a non-linear least-squares fit
yields $\omega_{p,D} = 8630$~cm$^{-1}$ and $1/\tau_D = 398$~cm$^{-1}$ at 295~K.
The observed value of $\sigma_{dc} \equiv \sigma_1(\omega \rightarrow 0)
\simeq 3100$~$\Omega^{-1}$cm$^{-1}$ is in reasonable agreement with transport
measurements.\cite{wang09,tanatar09}  As the temperature is reduced $\omega_{p,D}$
remains relatively constant, but the scattering rate decreases to $1/\tau_D =
285$~cm$^{-1}$ at 150~K.
The character of the conductivity changes dramatically below $T_N$.  The Drude
component weakens and narrows, with $\omega_{p,D} = 3970$~cm$^{-1}$ and
$1/\tau_D = 39$~cm$^{-1}$ at 80~K, resulting in a loss of spectral weight
that appears to be transferred to midinfrared band at $\simeq 1000$~cm$^{-1}$.
[The spectral weight is defined simply as the weight under the optical conductivity
curve over a given interval, $\int_{0^+}^{\omega_c} \sigma_1(\omega,T) d\omega$.]
%
%
Despite the nearly 80\% reduction in the number of free carriers, the
resistivity continues to decrease due to the dramatic reduction in the
scattering rate.    These observations are consistent with those of a
previous study;\cite{hu08} however, they are not the main focus of
this work.  Instead, we note in Fig.~\ref{fig:sigma} that in addition to the
broad features associated with the optical conductivity, there are two very
sharp resonances observed at $\simeq 94$ and 253~cm$^{-1}$ at 295~K.  The
vibrational features in the optical conductivity have been fit using Lorentz
oscillators with a linear background and the results are shown in
Table~\ref{tab:fits} at 295 and 6~K.

%
%

%
%
For $T > T_N$, BaFe$_2$As$_2$ is in the tetragonal $I4/mmm$ space group.  The
irreducible vibrational representation in the high-temperature tetragonal
(HTT) phase is\cite{litvinchuk08}
%
%
$$
  \Gamma_{vib}^{\rm HTT} = A_{1g} + B_{1g} + 2E_g + 2A_{2u} + 2E_u.
$$
Of these, only the $A_{2u}$ and $E_u$ vibrations are infrared active (along
the {\em c} axis and {\em a-b} planes, respectively), so the two modes we observe
are the symmetry-allowed infrared-active $E_u$ modes.  For $T < T_N$ the material
is in the orthorhombic $Fmmm$ space group, and the irreducible vibrational
representation of the low-temperature orthorhombic (LTO) phase is
$$
  \Gamma_{vib}^{\rm LTO} = A_g + B_{1g} + 2B_{2g} + 2B_{3g} + 2B_{1u} + 2B_{2u} + 2B_{3u}.
$$
The $B_{1u}$ modes are active along the {\em c} axis, and the orthorhombic
distortion lifts the degeneracy of the $E_u$ mode and splits it into $B_{2u} +
B_{3u}$ (active along the {\em b} and {\em a} axes, respectively) for a
total of four infrared-active modes at low temperature.  However,
{\em ab initio} studies indicate that the splitting of the $E_u$ mode in
the related LaFeAsO compound should be quite small,\cite{yildirim08} of the
order of $1.5$~cm$^{-1}$ (0.2~meV), and indeed no new modes
are observed in that material at low temperature.

Below $T_N$
at 6~K, the low-frequency mode has hardened somewhat to 95.4~cm$^{-1}$
and is now somewhat broader, suggesting that this mode may be showing signs of
splitting; however, the oscillator strength has not changed appreciably.  This
mode involves displacements primarily of the Ba atoms.\cite{litvinchuk08}
%
%
\begin{table}[tb]
\caption{The vibrational parameters for oscillator fits to the symmetry-allowed
infrared-active phonon modes observed in the {\em a-b} plane at 295 and 6~K,
where $\omega_j$, $\gamma_j$ and $\Omega_j$ are the frequency, width and
oscillator strength of the $j$th mode.  The estimated errors are indicated in
parenthesis. All units are in cm$^{-1}$.}
\begin{ruledtabular}
\begin{tabular}{ccc ccc}
 \multicolumn{3}{c}{295~K (HTT)} & \multicolumn{3}{c}{6~K (LTO)} \\
  $\omega_j$ & $\gamma_j$ & $\Omega_j$ &
  $\omega_j$ & $\gamma_j$ & $\Omega_j$ \\
  \cline{1-3} \cline{4-6}
%
%
  $94.0\,(0.1)$ & $3.5\,(0.4)$ & $222\,(8)$ &  $95.4\,(0.2)$ & $3.9\,(0.9)$ & $236\,(17)$ \\
 $253.2\,(0.1)$ & $4.1\,(0.2)$ & $226\,(7)$ &  $257.5\,(0.1)$ & $2.7\,(0.2)$ & $315\,(8)$ \\
\end{tabular}
\end{ruledtabular}
\label{tab:fits}
\vspace*{-0.5cm}
\end{table}

\begin{figure}[tb]
%
%
\vspace*{0.2cm}%
\centerline{\includegraphics[width=2.5in]{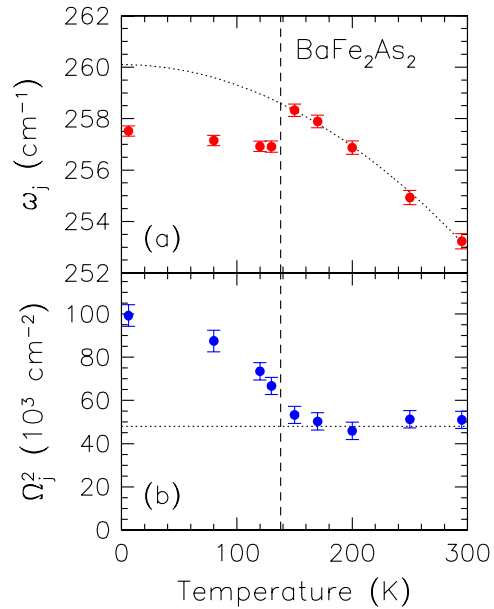}}%
\vspace*{-0.2cm}%
\caption{(Color online). The temperature dependence of the (a) frequency
  ($\omega_j$) and (b) intensity ($\Omega_j^2$) of the infrared-active mode
  in BaFe$_2$As$_2$ observed at $\simeq 253$~cm$^{-1}$.  In both cases the
  dashed line indicates $T_N \simeq 138$~K, and the dotted line represents
  the temperature dependence expected in the absence of a structural or
  magnetic transition.}%
\vspace*{-0.5cm}
\label{fig:phonon}
\end{figure}

%
%
The behavior of the 253~cm$^{-1}$ mode is fundamentally different.  Between 295
and 6~K this mode increases slightly in frequency and narrows
slightly (as expected); however, the oscillator strength increases from $\Omega_j =
226 \rightarrow 315$~cm$^{-1}$, leading to a doubling in the intensity (inset of
Fig.~\ref{fig:sigma}).  The detailed temperature dependence of the frequency and
the intensity of this mode is shown in Fig.~\ref{fig:phonon}.
In Fig.~\ref{fig:phonon}(a), the mode increases in frequency with decreasing temperature,
but at $T_N$ there is an abrupt decrease in frequency; for $T < T_N$ the position of
the mode displays little temperature dependence.  In Fig.~\ref{fig:phonon}(b) the intensity
remains constant from 295 to 150~K, but increases in a mean-field way for $T < T_N$, nearly
doubling in intensity at low temperature.  The dotted lines in Fig.~\ref{fig:phonon} represent
the expected behavior; the frequency of a mode is generally expected to follow a quadratic
temperature dependence (``hardening'').
%
%
The intensity of an infrared-active mode is related to the net dipole moment
$\mu_i = \sum_j Z_j^\ast u_{ij}$, where $Z_j^\ast$ is the Born effective charge of
the $j$th atom in the unit cell, and $u_{ij}$ is its displacement in the $i$th
direction;\cite{dowty87} the intensity of a mode is proportional to $\sum_i \mu_i^2$.
From this expression, the intensity of a mode is expected to remain constant,
unless there is a change in bonding or coordination.  Alternatively, the intensity
may also change if the electronic screening decreases, or if the lattice mode couples
to either the spins or the electronic background.  We will begin with the last point first.
%
%
The interaction of a lattice mode with the electronic background often results
in interference effects resulting in an asymmetric line shape;\cite{fano61,bozio87}
however, the observed phonon line shape in the optical conductivity is a symmetric
Lorentzian (inset of Fig.~\ref{fig:sigma}), suggesting that any coupling between
the lattice mode and the electronic background is small.  Spin-phonon coupling
has been observed in some quantum magnets to manifest itself primarily as a weak softening
of a phonon mode;\cite{choi03} however, there is virtually no effect on the intensity,
suggesting this type of coupling is also rather weak.
%
%
%

%
%
%
A decrease in the electronic background may lead to a reduction in the screening
and an increase in the oscillator strength.  However, in many ``bad metals''
screening effects are quite small.\cite{homes00}  While there is a substantial
decrease in the electronic background beneath both the 94 and 253~cm$^{-1}$ modes
below $T_N$, the strength of the 253~cm$^{-1}$ mode increases dramatically while
the strength of the 94~cm$^{-1}$ mode remains essentially unchanged.  In addition,
the electronic background beneath the 253~cm$^{-1}$ mode actually increases between
295 and 150~K, yet no change in intensity is observed.  Finally, in
the potassium-doped compound this mode may still be observed;\cite{li08} however,
in the cobalt-doped analog\cite{tu09} this mode is either not observed or extremely
weak, despite the fact that the electronic backgrounds are similar.  This evidence
suggests that the 253~cm$^{-1}$ mode is probably unscreened, but very sensitive
to local disorder in the Fe-As layers, which is not surprising given that this
vibration involves displacements of the Fe and As atoms.\cite{zhao09}

%
%
This brings us to the possibility of changes to the bonding or coordination. The structural
distortion in this material is rather weak and does not result in any significant changes
in coordination for the atoms in the unit cell.\cite{rotter08a}  It is possible that below
$T_N$ there might be a redistribution of charge.\cite{choi08}  Changes in $Z^\ast_{\rm Fe}$,
$Z^\ast_{\rm As}$, or both, would likely produce a change in $Z^\ast_{\rm Ba}$; however, the
intensity of the Ba mode at 94~cm$^{-1}$ does not change appreciably below $T_N$, suggesting
that $Z^\ast_{\rm Ba}$ is relatively constant.  This makes it unlikely that the increase in
intensity of the 253~cm$^{-1}$ mode results from a redistribution of charge.
Recent experimental and theoretical studies on the electronic structure of BaFe$_2$As$_2$
and related systems\cite{shimojima09,chen09,phillips09,lee09} conclude that the magnetism and the
structural distortion is driven by hybridization and orbital ordering.  In this treatment,
the structural distortion and magnetic order result from a hybridization of the four-fold
coordinated Fe $3d$ and the tetrahedrally positioned As $4p$ orbitals which strongly
modifies the tails of the Wannier functions (local real-space orbitals) perpendicular to
their original directions, resulting in a rare ferro-orbital ordering.\cite{lee09}
The change in the nature of the bonding between the Fe and As atoms implies that the
atomic displacements may be altered in a fashion that would lead to an increase in the
intensity of the 253~cm$^{-1}$ mode; this mechanism may also explain the abrupt shift
in the frequency at $T_N$.  However, the full extent of the effects of the orbital
ordering on the frequency and strength of this infrared-active mode will have to
wait for a more detailed calculation.\cite{lee09}

%
%
In summary, the detailed optical properties of BaFe$_2$As$_2$ have been determined
above and below $T_N \simeq 138$~K.  We have identified both symmetry-allowed
infrared-active $E_u$ modes at $\sim 94$ and 253~cm$^{-1}$ at 295~K.  In agreement
with earlier work, we observe a loss of spectral weight in the Drude component below
$T_N$ corresponding to an almost 80\% decrease in the number of free carriers;
this spectral weight is transferred to a midinfrared band.
In addition, we note the anomalous behavior of the 253~cm$^{-1}$ mode which undergoes
a discontinuous shift in frequency at $T_N$, and which doubles in intensity for
$T \ll T_N$.  While there are several possible mechanisms by which this increase in
intensity might be achieved, it is likely that a change in the nature of the bonding
between the Fe and As atoms due to orbital ordering below $T_N$ alters the character
of the atomic displacements, resulting in an increase in the net dipole moment.

%
%
We would like to acknowledge useful discussions with W. Ku, C.-C. Lee,
M. Strongin, and W.-G. Yin.
This work was supported by the National Science Foundation of China, the
National Basic Research Program of China (Nos. 2006CB601003 and 2007CB925001)
and the PCSIRT project of the Ministry of Education of China (IRT0754).
Work at BNL is supported by the Office of Science, U.S. Department of Energy
(DOE) under Contract No. DE-AC02-98CH10886.

%
%
%
\bibliography{bafeas}

\end{document}